\documentclass[11pt,a4paper]{article}
\usepackage{amsmath}
\usepackage{float}
\usepackage{caption2}
\usepackage{graphicx}

\newcommand{\ra}{\ensuremath {\rangle} }

\begin{document}

\title{\textbf{\large{Generalization of the Deutsch algorithm using two qudits}}}
\author{Jos\'{e} L.\ Cereceda\thanks{Electronic mail: jl.cereceda@telefonica.net} \\
\textit{C/Alto del Le\'{o}n 8, 4A, 28038 Madrid, Spain}}

\date{\today}

\maketitle

\begin{abstract}
Deutsch's algorithm for two qubits (one control qubit plus one auxiliary qubit) is extended to two $d$-dimensional quantum systems or qudits for the case in which $d$ is equal to $2^n$, $n=1,2,\ldots$ . This allows one to classify a certain oracle function by just one query, instead of the $2^{n-1}+1$ queries required by classical means. The given algorithm for two qudits also solves efficiently the Bernstein-Vazirani problem. Entanglement does not occur at any step of the computation.

\vspace{.2cm}


\end{abstract}

\vspace{.4cm}

In the original Deutsch's problem \cite{Deutsch85} the task is to ascertain a global property of an unknown but fixed Boolean function $f: \{0,1\} \to \{0,1\}$, namely whether the function is constant (i.e. $f(0)\oplus f(1) = 0$) or balanced (i.e. $f(0)\oplus f(1) = 1$), where $\oplus$ denotes addition modulo 2. Suppose we are given a ``black box'', or oracle, computing the function. Classically, it is obvious that one has to
evaluate both $f(0)$ and $f(1)$ to solve the problem (that is, one needs to perform two function calls or queries). By allowing linear superpositions of single qubit states, however, Deutsch \cite{Deutsch85} devised a quantum algorithm which, with probability $50\%$, provides the correct answer for $f$ (namely, whether it is constant or balanced) with a single evaluation of the function. An improved, deterministic version of Deutsch's algorithm has been given by Cleve et al. \cite{CEMM98} which provides the exact answer for $f$ in all cases. The speedup of the quantum algorithm over the classical one in this case is thus a factor of two. Deutsch's problem was generalized by Deutsch and Jozsa (DJ) \cite{DJ92} to cover Boolean functions of the type $f: \{0,1\}^{n} \to \{0,1\}$. The task is again to determine whether $f$ is constant or balanced, where balanced means that $f$ yields the value $0$ for exactly half of the arguments and $1$ for the rest. To solve this problem classically, it is necessary to get the function evaluated for $2^{n-1}+1$ arguments in the worst case. Quantum mechanically, a single evaluation of $f$ is sufficient \cite{CEMM98,DJ92}. The exponential speedup achieved by the DJ algorithm stems primarily from the fact that it allows for quantum superpositions involving an exponentially large number $2^n$ of orthogonal states of $n$ qubits.

Another possible way to enhance the power of quantum computing is to increase the dimensionality of the individual quantum systems involved in the computation. In this paper we provide an algorithm using two $d$-dimensional quantum systems or qudits (one control qudit plus one auxiliary qudit) of dimension $d=2^n$, which solves the generalized Deutsch's problem with a single function evaluation. Note that the DJ algorithm employs $n+1$ qubits ($n$ control qubits plus one auxiliary qubit) to solve the same problem \cite{CEMM98,DJ92}. As both quantum algorithms (DJ's and ours) accomplish their goal with a single function evaluation we deduce that the computational capacity of a single qudit of dimension $d=2^n$ is exactly the same as that corresponding to $n$ qubits. Of course, this is directly related to the fact that the dimension of the state space of a single qudit with $d=2^n$, is the same as the dimension of the state space associated with $n$ qubits. The advantage of our algorithm resides in the fact that the control register consists of a \textit{single\/} quantum system. Indeed it is important to emphasize from the start that, in our algorithm, entanglement does not take place at any stage of the computation. This contrasts with the DJ algorithm where, in general, entanglement does indeed occur between the qubits of the control register \cite{J+EJ,CKH98,ABV01}. Moreover, for a register consisting of a single $d$-level quantum system, the steps of preparation, unitary evolution, and measurement involved in the quantum computation, should, at least for enough low values of $d$, be easier to perform than it would for the case of $n$ distinct qubits. To substantiate this view, we must refer to previous work by Kessel and Ermakov \cite{KE991,KE992,KE001,KE002} (see also Ref.~\cite{KY02}). These authors, using the virtual-spin formalism \cite{KE991,KE992}, showed how the four states of a single quantum particle (specifically, a spin-3/2 nucleus) can be used to write and read two quantum bits of information, to prepare the initial state, and to implement a full set of two-qubit gates \cite{KE991,KE992,KE001}. Analogously, they showed that a quantum particle with eight energy levels can store three qubits, and that three-qubit gates can be physically realized on one spin-7/2 particle \cite{KE001,KE002}. On the other hand, Ahn et al. \cite{Ahn00} experimentally investigated the storage and retrieval of information in the quantum phase of a coherent superposition state of energy levels in an $N$-state Rydberg atom, and, in particular, they demonstrated storage of numbers up to $2^{N-1}$ for $N=8$. Furthermore, Ahn et al. \cite{Ahn00} suggested that a straightforward extrapolation of their results would allow numbers as large as $2^{100}$ to be stored in a single $N$-level atom, with $N=20$. It should be noticed, however, that, although algorithms using single quantum particle as control register may not require entanglement at any step of the computation, a physical realization of them would surely require an exponential overhead in some other recourse(s) for high values of $d$ \cite{J+EJ,Lloyd99,Meyer00}.

Let us now define more precisely the problem under consideration. Let
\begin{equation}
f: \{0,1,\ldots,d-1\} \to \{0,1\}
\end{equation}
be a given function that maps each of the $d$ arguments $0,1,\ldots,d-1$ to a one-bit value, with $d$ being an even number $2^n$, and $n=1,2,\ldots$ . The function is constrained to be either constant or balanced, i.e. it fulfills the property that
\begin{equation}
f(0)\oplus f(1)\oplus\cdots\oplus f(d-1) = 0,
\end{equation}
for a constant function, or
\begin{equation}
f(0)\oplus f(1)\oplus\cdots\oplus f(d-1) = d/2,
\end{equation}
for a balanced one. Unless stated otherwise, in Eqs.~(2) and (3), and in the remainder of this paper, $\oplus$ denotes addition modulo $d$. Given an oracle that evaluates the function for a given argument, our problem is again to decide, by queries to the oracle, whether $f$ is constant or balanced. Note that the case of $d=2$ corresponds to the problem originally considered by Deutsch \cite{Deutsch85}. Any (deterministic) classical algorithm for this problem would, in the worst case scenario, require $d/2 +1$ function calls to know with certainty which kind of function we have at hand. As we shall presently see, there is a quantum algorithm involving two qudits that solves this problem with a single evaluation of $f$. This algorithm uses a quantum gate $U_f$ for two qudits that is a direct generalization of the $f$-controlled-NOT gate for two qubits used in the Deutsch algorithm. Let us denote the set of computational basis states of a qudit by $\{|0\ra,|1\ra,\ldots,|d-1\ra\}$. The operation of the two-qudit gate $U_f$ is completely defined by its action on the computational basis for each qudit:
\begin{equation}
|x\ra |y\ra \stackrel{U_f}{\longrightarrow} |x\ra |y\oplus f(x)\ra,
\end{equation}
where $|x\ra$ and $|y\ra \in \{|0\ra,|1\ra,\ldots,|d-1\ra\}$ denote the state of the control and auxiliary qudits, respectively. We  may call the operation performed by the $U_f$ gate a $f$-controlled-SHIFT operation, for the effect of $U_f$ on the auxiliary qudit when $f(x)=1$ is to ``shift'' its state from $|y\ra$ to the adjacent state $|y\oplus 1\ra$ ($y=0,1,\ldots,d-1$).

\setlength{\unitlength}{1cm}

\begin{figure}[ttt]
\begin{picture}(12,4)(-2.7,0.25)
\thicklines

\put(1.5,1.1){\large{$|1\ra$}}
\put(2.1,1.2){\vector(1,0){1.0}}
\put(3.1,0.8){\framebox(0.8,0.8){\large\textit{\textbf{H}}}}
\put(3.9,1.2){\line(1,0){0.8}}\put(4.7,0.8){\mathversion{bold}\framebox(0.8,0.8){\large{$U_f$}}}
\put(5.5,1.2){\vector(1,0){1.1}}\put(6.8,1.1){\large{$|1_H\ra$}}

\put(1.5,2.8){\large{$|0\ra$}}
\put(2.1,2.9){\vector(1,0){1.0}}
\put(3.1,2.5){\framebox(0.8,0.8){\large\textit{\textbf{H}}}}
\put(3.9,2.9){\line(1,0){1.2}}\put(5.1,2.9){\circle*{0.3}}\put(5.1,2.9){\line(0,-1){1.3}}
\put(5.1,2.9){\line(1,0){1.2}}\put(6.3,2.5){\framebox(0.8,0.8){\large\textit{\textbf{H}}}}
\put(7.1,2.9){\line(1,0){1}}
\put(8.1,1.9){\framebox(0.8,2.0){\large\textit{\textbf{M}}}}

\put(8.9,3.7){\line(1,0){0.7}}\put(9.8,3.6){\large{$|0\ra$}}
\put(8.9,3.3){\line(1,0){0.7}}\put(9.8,3.2){\large{$|1\ra$}}
\put(8.9,2.9){\line(1,0){0.7}}\put(9.8,2.8){\large{$|2\ra$}}
\put(9.2,2.35){\large{$\vdots$}}
\put(8.9,2.1){\line(1,0){0.7}}\put(9.8,2.0){\large{$|d\!-\!1\ra$}}
\end{picture}

\vspace{-.4cm}

\renewcommand{\figurename}{\small{Fig.}}
\setcaptionmargin{1.2cm}
\caption{\small{The network diagram for the generalized Deutsch's algorithm using two qudits. A final measurement result compatible with the control qudit being in the state $|0\ra$ means that the function was constant. Any other output $|x\ra$ (with $x=1,2,\ldots,d-1$) means that the function was balanced.}}

\end{figure}
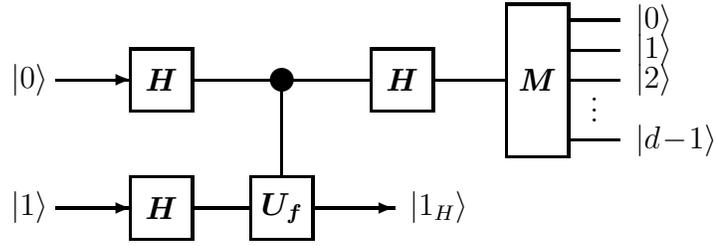

The actual circuit implementing our algorithm is shown in Fig.~1. The initial state of the qudits in the quantum network is $|0\ra|1\ra$, where the first (second) ket always refers to the state of the control (auxiliary) qudit. In the first step, both qudits undergo a unitary transformation $H_d$ (a qudit Hadamard gate) such that the state of the control qudit is transformed to
\begin{equation}
|0\ra \to |0_H\ra = |0\ra + |1\ra +\cdots + |d-1\ra ,
\end{equation}
(apart from an unimportant normalization factor, which will be omitted in the following), whereas the state of the auxiliary qudit is transformed to
\begin{equation}
|1\ra \to |1_H\ra = |0\ra - |1\ra +\cdots + |d-2\ra - |d-1\ra ,
\end{equation}
with the $+$ and $-$ signs following each other alternately through the superposition. Then the state after the first two Hadamard gates is
\begin{equation}
|0_H\ra |1_H\ra = \sum_{x,y=0}^{d-1} (-1)^{y} |x\ra |y\ra.
\end{equation}
Next, we apply the unitary gate $U_f$ on this state (the middle operation shown in Fig.~1). Of course, the actual computation of the function $f$ comes from the action of this $f$-controlled-SHIFT gate. To determine the effect of $U_f$ on the state (7), first note that, for each $x=0,1,\ldots,d-1$, we have
\begin{equation}
U_f \left( \sum_{y=0}^{d-1} (-1)^{y}|x\ra |y\ra \right) = \sum_{y=0}^{d-1} (-1)^{y} |x\ra |y\oplus f(x)\ra =
(-1)^{f(x)} |x\ra |1_H\ra,
\end{equation}
since $f(x)=0$ or $1$. The rightmost expression of (8) follows at once from the fact that, actually, the auxiliary qudit state $|1_H\ra$ is an eigenstate with eigenvalue $(-1)^{f(x)}$ of the operator that sends the state $|y\ra$ to $|y\oplus f(x)\ra$. Therefore, the total state after the $f$-controlled-SHIFT is
\begin{equation}
\left( \sum_{x=0}^{d-1} (-1)^{f(x)} |x\ra \right) |1_H\ra \equiv |\chi_f \ra |1_H\ra .
\end{equation}

From (9), we see that the state of the auxiliary qudit remains unchanged while each component $|x\ra$ of the control qudit acquires a phase factor of $(-1)^{f(x)}$. Clearly, for a constant function, the state $|\chi_f\ra$ is simply $|0_H\ra$. On the other hand, for a balanced function, the resulting state $|\chi_f\ra$ will always be \textit{orthogonal\/} to the state $|0_H\ra$, since now $|\chi_f\ra$ consists of an equally weighted superposition with exactly half of the $|x\ra$'s having a minus sign. Therefore, the two possibilities (namely, constant $f$ or balanced $f$) can be reliably distinguished by means of a projective measurement $P = |0_H\ra \langle 0_H |$ on the control qudit. So, if $f$ is constant (respectively, balanced), the probability of observing the control qudit in the state $|0_H\ra$ is 1 (0), so that a measurement filtering the state $|0_H\ra$ will with certainty give the result ``1'' (``0''). Alternatively, we may first apply a Hadamard transformation to the state $|\chi_f\ra$ in order to unitarily rotate the eigenbasis of the measurement into the computational basis $\{|0\ra,|1\ra,\ldots,|d-1\ra\}$. Thus, for constant $f$, the state of the control qudit after the last Hadamard gate becomes $|0\ra$, and for balanced $f$ the resulting state becomes orthogonal to this, so that a subsequent measurement of the control qudit in the computational basis will distinguish these cases with certainty. (This is what is indicated in the last step of the diagram by the symbol $M$, Fig.~1.) As an example, for $d=4$, the Hadamard operation has the possible matrix representation
\begin{equation}
H_4 = \frac{1}{2} \left( \begin{array}{rrrr}
1 & 1 & 1 & 1 \\
1 & -1 & 1 & -1 \\
1 & 1 & -1 & -1 \\
1 & -1 & -1 & 1
\end{array}  \right).
\end{equation}
Note that $H_{4}^{2} = I_4$, and then, for example, we have that $H_4 |0\ra = |0\ra + |1\ra +|2\ra + |3\ra$ and, conversely, $H_4 (|0\ra + |1\ra +|2\ra + |3\ra) = |0\ra$. Similarly, we have that $H_4 |1\ra = |0\ra - |1\ra +|2\ra - |3\ra$, and $H_4 (|0\ra - |1\ra +|2\ra - |3\ra) = |1\ra$. Analogously, for $d=8$, the Hadamard transformation matrix has the possible form
\begin{equation}
H_8 = \frac{1}{2\sqrt{2}} \left( \begin{array}{rrrrrrrr}
1 & 1 & 1 & 1 & 1 & 1 & 1 & 1 \\
1 & -1 & 1 & -1 & 1 & -1 & 1 & -1 \\
1 & 1 & -1 & -1 & 1 & 1 & -1 & -1 \\
1 & -1 & -1 & 1 & 1 & -1 & -1 & 1 \\
1 & 1 & 1 & 1 & -1 & -1 & -1 & -1 \\
1 & -1 & 1 & -1 & -1 & 1 & -1 & 1 \\
1 & 1 & -1 & -1 & -1 & -1 & 1 & 1 \\
1 & -1 & -1 & 1 & -1 & 1 & 1 & -1 
\end{array}  \right).
\end{equation}
Note again that $H_{8}^{2} = I_8$, so that $H_8 |x\ra = |x_H\ra$ and, conversely, $H_8 |x_H\ra = |x\ra$, for each $x=0,1,\ldots,7$. In general, the Hadamard operator acting on a single qudit of dimension $d=2^n$ is defined as
\begin{equation}
H_d |x\ra = \frac{1}{\sqrt{d}}\sum_{x^{\prime}=0}^{d-1} (-1)^{x \cdot x^{\prime}} |x^{\prime}\ra,
\end{equation}
where
\begin{equation}
x \cdot x^{\prime} = \oplus_{i=0}^{n-1} \, x_i x_i^{\prime}
\end{equation}
denotes the bitwise inner product of $x$ and $x^{\prime}$ expressed in the binary representation $x= \sum_{i=0}^{n-1} x_i 2^{i}$ and $x^{\prime}= \sum_{i=0}^{n-1} x_i^{\prime} 2^{i}$, with $x_{i},x_{i}^{\prime} \in \{0,1\}$. Please note that the symbol $\oplus$ in Eq.~(13) denotes addition modulo 2. It is easily seen that, for each computational basis state $|x\ra$, $H_d (H_d |x\ra) = |x\ra$, so that the operator $H_d$ indeed fulfills $H_d^2 = I_d$. We further note that the matrix $H_{2^{n}}$ can be obtained from the matrix $H_2 = \frac{1}{\sqrt{2}} \left( \begin{smallmatrix} 1&1 \\ 1&-1 \end{smallmatrix}\right)$ by performing $n$ times the tensor product of $H_2$ by itself, i.e. $H_{2^{n}} = \otimes^{n} H_2$.

Provided with the general expression for the Hadamard transformation, Eq.~(12), it is immediate to show that the quantum network in Fig.~1 can equally be used to solve the Bernstein-Vazirani (BV) problem \cite{BV93} (see also Ref.~\cite{CEMM98}): given an oracle which evaluates for some $n$-bit string $a$, the function
\begin{equation}
f_a (x) = a \cdot x = \oplus_{i=0}^{n-1} \, a_i x_i
\end{equation}
of domain $x \in \{0,1\}^{n}$, the task is to determine $a$ by queries to the oracle. Note that there is a one-to-one correspondence between the $n$-bit domain $\{0,1\}^n$ and its decimal representation $\{0,1,\ldots,d-1\}$, so that the function $f_a$ can equivalently be thought of as mapping the domain $\{0,1,\ldots,d-1\}$ to a one-bit value, as long as $d=2^n$. Classically, to determine $a$ it is necessary to perform at least $n$ function evaluations, since $a$ contains $n$ bits of information and each classical function evaluation returns a single bit of information \cite{CEMM98}. However, if the circuit in Fig.~1 is used with the function $f_a$, then $a$ can be determined with a single function evaluation. Indeed, in this case, and for the same initial state $|0\ra |1\ra$ as above, Eq.~(9) reads
\begin{equation}
\left( \sum_{x=0}^{d-1} (-1)^{a \cdot x} |x\ra \right) |1_H\ra \equiv |\chi_a \ra |1_H\ra .
\end{equation}
Comparing the expression in parenthesis of (15) with (12), we can see that the control qudit state after the $f_a$-controlled-SHIFT is simply the Hadamard transformation of the state $|a\ra$, i.e. $|\chi_a \ra = H_d |a\ra$. Therefore, after the action of the last Hadamard gate of the circuit in Fig.~1, the state $|\chi_a \ra$ becomes $H_d |\chi_a \ra = H_{d}^{2} |a\ra = |a\ra$. Thus, a final measurement of the control qudit in the computational basis will yield with certainty the integer $a$.

Summing up, we have described a quantum algorithm using two qudits (one control qudit plus one auxiliary qudit) of dimension $d=2^n$ that solves with certainty the generalized Deutsch problem, namely the problem of determining whether the function $f: \{0,1,\ldots,d-1\} \to \{0,1\}$ is constant or balanced, with a single function evaluation followed by a measurement of the state of the control qudit. If we measure directly in the non-computational basis $\{|0_H\ra,|1_H\ra,\ldots,|(d-\nolinebreak 1)_H\ra\}$, then finding the control qudit in the state $|0_H\ra$ unambiguously signals that the function was constant. Otherwise, if the control qudit is found in a state orthogonal to $|0_H\ra$, then $f$ was balanced. On the other hand, if we measure in the computational basis $\{|0\ra,|1\ra,\ldots,|d-\nolinebreak 1\ra\}$ (after applying a Hadamard transformation to the control qudit) and obtain the outcome corresponding to $|0\ra$, then the function was constant. For any other output the function was balanced. Furthermore, we have shown that the same quantum algorithm using two qudits can be used to solve efficiently the BV problem.

An important feature of our algorithm is that entanglement is never present throughout the quantum computation. Indeed, from Eq. (9), it is apparent that, after the action of the $f$-controlled-SHIFT gate, the control and auxiliary qudit states $|\chi_f\ra$ and $|1_H\ra$ are unentangled from each other. Furthermore, each one of these states corresponds to a single quantum system and, therefore, entanglement is not an issue in the present algorithm.\footnote{
\small{It is worth pointing out that the BV algorithm does not require entanglement even for the case where the control register consists of $n$ distinct qubits \cite{Meyer00}. Indeed, in this case, and for any $n$-bit string $a$, the expression in parenthesis of (15) can be written as a tensor product of $n$ single qubit states. This factorization is a direct consequence of the above mentioned relation $H_{2^{n}} = \otimes^{n} H_2$.} This contrasts with the DJ algorithm for $n$ control qubits, where entanglement is in general necessary \cite{J+EJ,CKH98,ABV01}.}
Moreover, it should be noticed that our algorithm for two qudits can work equally well if the auxiliary qudit is replaced by a single qubit. In this case the gate $U_f$ would correspond to a $f$-controlled-NOT gate defined by Eq.~(4) (where now $|x\ra \in \{|0\ra,|1\ra,\ldots,|d-1\ra\}$ and $|y\ra \in \{|0\ra,|1\ra\}$), and the qubit state $|1_H\ra$ would be simply $|0\ra - |1\ra$, so that the joint (qudit plus qubit) state after the $U_f$ gate would be given similarly by Eq.~(9). Now, as the auxiliary qubit in state $|1_H\ra$ is not altered during the computation, we might further refine our algorithm by eliminating the auxiliary qubit while retaining the control qudit \cite{CKH98}. In this case, the function evaluation can be carried out by means of the $f$-controlled gate, whose action on the (computational) basis elements $|x\ra$ of the control qudit is defined as
\begin{equation}
U_f |x\ra = (-1)^{f(x)} |x\ra.
\end{equation}
Note that, when this refinement is made, we are indeed dealing with a single particle quantum computer!

To conclude, it will be added that the logic behind the two-qudit gate $U_f$ can be naturally extended to deal with multivalued functions of the type
\begin{equation}
f: \{0,1,\ldots,d-1\} \to \{0,1,\ldots,d-1\} ,
\end{equation}
in such a way that the operation of the $U_f$ gate is again defined by Eq.~(4), but now we let $f$ take the values $f(x)=0,1,\ldots,d-1$. It is readily seen that the effect of this generalized, multivalued gate on the state $|0_H\ra |1_H\ra$ is given by
\begin{equation}
U_f |0_H\ra |1_H\ra = |\chi_f \ra |1_H\ra ,
\end{equation}
where the transformed control qudit state $|\chi_f \ra$ is in turn given by
\begin{equation}
|\chi_f \ra  = \sum_{x=0}^{d-1} (-1)^{f(x)} |x\ra .
\end{equation}
Again, Eqs.~(18) and (19) are a consequence of the fact that $|1_H\ra$ is an eigenstate with eigenvalue $(-1)^{f(x)}$ of the operator that sends $|y\ra$ to $|y\oplus f(x)\ra$, $f(x)=0,1,\ldots,d-1$. Now we define the following two classes of functions. We say that a function $f$ has a constant parity when all values $f(0),f(1),\ldots,f(d-1)$ are either even or odd. On the other hand, we say that $f$ has a balanced parity when it yields an even value for exactly half of the arguments, and an odd value for the other half. A concrete example of a function with constant (balanced) parity for $d=8$, is: $(f(0),\ldots,f(7)) = (4,2,0,0,0,6,2,4)$ (respectively, $(4,2,0,0,1,1,7,5)$). Then it is clear that if $f$ has constant parity the state $|\chi_f \ra$ is simply $|0_H\ra$, whereas if $f$ has balanced parity the state $|\chi_f \ra$ is orthogonal to $|0_H\ra$, since for this case exactly half of the terms in the superposition (19) have a minus sign. Thus, by suitably measuring the state of the control qudit, it can be determined with certainty whether $f$ has constant or balanced parity, thereby showing that our algorithm using two qudits can also solve efficiently the generalized Deutsch's problem for multivalued functions of the type (17). (Note that to solve this problem classically, $d/2+1$ function calls are required before determining the answer with certainty in the worst case scenario.)\footnote{
\small{For completeness, let us mention that by using a single $2^n$-level (respectively, $2^m$-level) quantum system as the control (auxiliary) register we can equally solve the parity problem for the more general function $f: \{0,1,\ldots,2^n-1\} \to \{0,1,\ldots,2^m-1\}$, where $n,m=1,2,\ldots$ .}}

Finally, one might wonder whether the higher computational power of qudits could be exploited to improve other existing quantum algorithms, or even to design entirely new ones. It is our hope that this will eventually be the case, and that the use of qudits and multivalued logic gates can serve as a valuable tool in the development of efficient quantum algorithms (see, in this respect, the work in Ref.~\cite{MS00} where the authors derive a set of one- and two-qudit gates that are sufficient for universal multivalued computing, and show how such gates can be implemented by using $d$-level ions in the linear ion trap model).

\end{document}